\documentclass[12pt,preprint]{emulateapj}
\def\gsim{\;\rlap{\lower 2.5pt
 \hbox{$\sim$}}\raise 1.5pt\hbox{$>$}\;}
\def\lsim{\;\rlap{\lower 2.5pt
   \hbox{$\sim$}}\raise 1.5pt\hbox{$<$}\;}

\begin{document}
\title{Simulations of XUV Disks with a Star Formation Density Threshold }
\author{Stephanie J. Bush\altaffilmark{1}, T.J. Cox\altaffilmark{1,2}, Lars Hernquist\altaffilmark{1},
   David Thilker\altaffilmark{3}, Joshua D. Younger\altaffilmark{1}}
\altaffiltext{1}{Harvard-Smithsonian Center for Astrophysics, 60 Garden St, Cambridge, MA 02143 USA}
\altaffiltext{2}{W.M. Keck Postdoctoral Fellow at the Harvard-Smithsonian Center for Astrophysics}
\altaffiltext{3}{Center for Astrophysical Sciences, The Johns Hopkins
  University, 3400 North Charles ST, Baltimore, MD 21218}

\email{sbush@cfa.harvard.edu}

\slugcomment{Accepted for publication in ApJL}

%\date{\today}

\begin{abstract}

The outer regions of disk galaxies show a drop-off in optical and H$\alpha$
emission, suggesting a star formation threshold radius, assumed to owe
to a critical surface density below which star formation does not take
place. Signs of filamentary star formation beyond this threshold radius have been observed in
individual galaxies in the H$\alpha$ and recent GALEX surveys have discovered that 30\% of disk
galaxies show UV emission out to  2-3 times the
optical radius of the galaxy.    We
run smooth particle hydrodynamics simulations of disk galaxies with
constant density extended gas disks to test whether over-densities owing
to spiral structure in the outer disk can reproduce the observed
star formation.  We indeed find that spiral density waves from the inner
disk propagate into the outer gas disk and raise local gas regions
above the star formation density threshold, yielding features similar
to those observed. Because the amount of star formation is low, we
expect to see little optical emission in outer disks, as observed.  Our results
indicate that XUV disks can be simulated simply by adding an extended gas
disk with a surface density near the threshold density to an isolated
galaxy and evolving it with fiducial star formation parameters.

\end{abstract}


\section{Introduction}

The most basic description of galactic-scale star formation is the
Schmidt relation, which assumes that
the star-formation rate (SFR) is a power law function of the gas density
\citep{Schmidt-1959}. \citet{Kennicutt-1998} showed that this relation
accurately describes star formation in galaxies over a range of $10^{5}$ in gas surface
density and $10^{6}$ in SFR per unit area.  However, beyond a critical galactocentric radius, SFRs determined from
H$\alpha$ emission are truncated
relative to the expectation from the Kennicutt-Schmidt relation despite the
availability of large reserves of low density cold gas \citep{Kennicutt-1989,
  Martin-Kennicutt-2001, Wong-Blitz-2002}. The lack of star formation at large
radii has primarily been attributed 
to dynamical disk stability, i.e., 
\begin{equation}
Q(r) = \frac{\Sigma_{c}(r)}{\Sigma_{gas}(r)} = \frac{\kappa(r) c_{s}}{\pi G
  \Sigma_{gas}(r)} \gsim 1 \, ,
\end{equation}
\noindent where $\kappa$ is the epicyclic frequency at a given radius,
$\Sigma_{c}(r)$ is the critical surface density and $c_{s}$ is sound speed of
the gas, proportional the
velocity dispersion of the gas ( $\sigma=c_{s}\gamma^{-1/2}$, where $\gamma$ is
the ratio of specific heats) \citep{Toomre-1964, Spitzer-1968, Quirk-1972}.
 The effective velocity dispersion of 
gas in spirals is roughly constant across
the disk \citep[e.g.,][]{vanderKruit-Shostak-1984, Anderson-et-al-2006}
leaving $\kappa$ as the only radially dependent quantity in the 
critical surface density. If velocity curves of spiral galaxies are
approximately flat, $\kappa$, and therefore $\Sigma_{c}$, fall with $r$,
implying that at some radius the disk becomes
gravitationally stable, which may inhibit star
formation.  In support of this notion, several studies have shown correlations between the radius at which
star formation ceases and the radius at which $Q\approx 2$ \citep[][see
  \citet{Schaye-2004} for a comparison of these results]{Kennicutt-1989,
  Martin-Kennicutt-2001, Wong-Blitz-2002}. More sophisticated models
including the formation of cold molecular gas  have also been
successful at reproducing observed SFR truncations \citep{
  Elmegreen-Parravano-1994, Schaye-2004} .

Observations indicate the presence of star formation beyond
the threshold. Faint, isolated
H\,II regions distributed in spiral structures out to two
optical radii have been found in face on galaxies \citep{Ferguson-et-al-1998}
and in cold disks in edge-on galaxies \citep{Christlein-Zaritsky-2008}. Similar star formation has been seen in the Milky
Way \citep{Brand-et-al-2001} and dwarf galaxies with gas densities below the
threshold \citep{vanZee-et-al-1997}. A systematic survey by the Galaxy Evolution Explorer
(GALEX: \citet{Martin-et-al-2005}) recently discovered that 30\% of
disk galaxies show extended UV emission beyond the optical radius, and presumed star formation
threshold, of the galaxy \citep{Thilker-et-al-2007, Gildepaz-et-al-2005,
  Thilker-et-al-2005, Zaritsky-Christlein-2007}. In 2/3 of these cases, this emission is
structured, lying in spiral or filamentary patterns (Type I XUV
disks). In the other 1/3, the emission takes the form of a large zone in the
outer regions of the galaxy with an enhanced 
UV/optical ratio relative to the inner disk
(Type II XUV disk) \citep[][hereafter T07]{Thilker-et-al-2007}.

To reconcile outer disk star formation with disk stability, it has been proposed that local densities above
$\Sigma_{c}(r)$ in the outer disk allow star formation beyond the
truncation radii \citep{Kennicutt-1989, Martin-Kennicutt-2001,
  Schaye-2004, Elmegreen-Hunter-2006, Gildepaz-et-al-2007}. The low
SFR implies a low number of O stars producing H$\alpha$
emission at a given time, or none at all, explaining why this star formation was
undetected in the H$\alpha$ in some galaxies \citep{Gildepaz-et-al-2007}. Most of these disks
have large amounts of H\,I in their outer regions and UV and H$\alpha$
complexes are often coincident with local H\,I over-densities 
\citep[][T07]{Ferguson-et-al-1998} supporting this
picture. Analysis of the star formation profiles
of these XUV disk galaxies indicates that the azimuthally averaged
Kennicutt-Schmidt relation could hold beyond the dynamical threshold
 radius \citep[][Hereafter B07]{Boissier-et-al-2007}. However, studies of star forming clumps
 in the XUV disk of M\,83 indicate that traditional star formation laws, including
the threshold radius, hold locally \citep{Dong-et-al-2008}.

Theoretically, \citet{Elmegreen-Hunter-2006} have shown that XUV disks may not be
inconsistent with a star formation density threshold.  In their analytical model, gas
clumping triggered by spiral density waves, radial variations in the interstellar
medium (ISM) turbulence, and gas phase transitions lead to local regions of star
formation in the extended gas disk.
In this paper, we investigate XUV disks using smooth particle hydrodynamic
(SPH) simulations of an isolated disk galaxy with an extended gas disk and a simple,
commonly employed, star formation prescription based on the Kennicutt-Schmidt relation
that includes a threshold density for star formation.
We demonstrate that, even with these fiducial star formation laws, spiral structure
propagates from the inner disk to the
extended H\,I disk and produces local regions of enhanced of gas density which
trigger star formation similar to that seen in observed Type I XUV disks.

\section{Method} \label{sec:method}

We used the SPH code \textsc{Gadget2} \citep{Springel-2005} to run
simulations of galaxies with extended gas disks. We incorporate a
sub-resolution multi-phase model of the ISM according to
\citet{Springel-Hernquist-2003}, which includes radiative cooling. The parameters of the multiphase model are
$t_{\star} = 6.5$ Gyr, $A_{0}= 3000$ and $ T_{SN} = 3\times10^{8}$. The
equation of state is determined by these parameters. No additional star
formation feedback is employed. Star formation from the
cold phase (total cold gas, our simulation does not treat atomic and molecular
gas separately) follows a
Schmidt volume density law $\rho_{SFR} \propto \rho_{gas}^{N}$ with $N=1.5$
that is normalized to give approximately the same SFR as the Milky Way. A local star
formation volume density cutoff of 0.004 M$_{\odot}$/pc$^{3}$ is included to
give a surface density threshold of $\sim$ 5-10
M$_{\odot}$/pc$^{2}$, consistent with observations \citep{Kennicutt-1989,
  Martin-Kennicutt-2001}.

\begin{figure}
\plotone{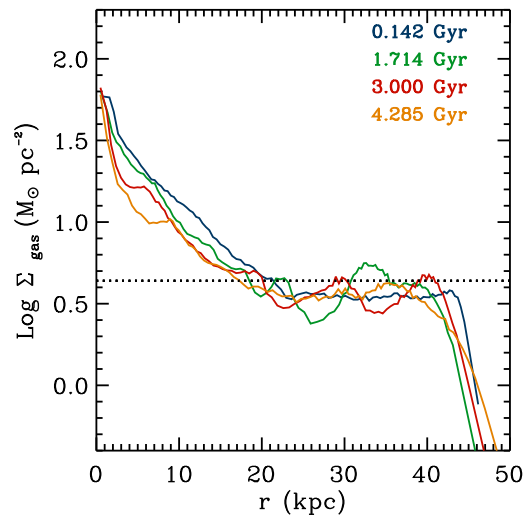}
\caption[]{Gas density profile of the simulated galaxy at the four
  different times shown in Figure~\ref{fig:sequence}. Notice how in the first
  simulation snapshot (0.142 Gyr, blue) the extended gas disk has a flat
  profile, while as the simulation goes on and spiral structure develops, the
  outer disk shows peaks that lie above the star formation threshold
  density. From the simulation, the volume density threshold  appears to
  correspond to a surface density threshold
  of approximately $\sim 4.5$ M$_{\odot}$/pc$^{2}$ which is shown by the
  dotted line.}
\label{fig:HIprofile}
\end{figure}

The progenitor galaxy models were constructed following
\citet{Springel-Dimatteo-Hernquist-2005}. The galaxy is roughly analogous
to the Milky Way with a total mass of $M_{200}\sim10^{12}$ M$_{\odot}$ composed
of a
\citet{Hernquist-1990} dark matter halo with $c=9$, a
stellar disk scale length of $3.7$ kpc (we use $h=0.72$ throughout this paper), an exponential gas disk of the
same scale length. However, we add an approximately 
constant density gas disk that extends to $45$ kpc and has a surface density
of $\sim 3.5$ M$_{\odot}$/pc$^{2}$. The baryonic mass fraction is
0.04. Forty percent of the baryons are in the stellar disk, 20\% are in the
exponential gas disk and 40\% are in the constant density gas disk.
The gas profile of the galaxy at four times in the
simulation is shown
in Figure~\ref{fig:HIprofile}. The galaxy has no bulge and a 0.01\% $M_{200}$ 
mass black hole in its center. The black hole is represented by a sink
particle that can accrete mass at the Bondi  rate
\citep{Springel-Dimatteo-Hernquist-2005}, but as we are interested in the
outer disk its presence does not affect our results. The models have $N_{halo} = 1.2\times 10^{5}$, $N_{stars}
= 8 \times 10^{4}$ and $N_{gas} = 1.6 \times 10^5$ particles in each
component.  The \textsc{Gadget} softening length of the baryonic particles and halo particles
are  0.14 kpc and 0.28 kpc, respectively.

\section{Results} \label{sec:results}

\begin{figure}
\plotone{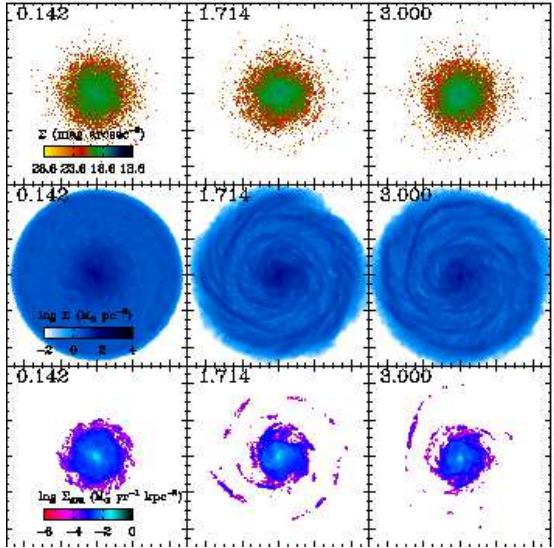}
\caption[]{Three snapshots from our simulation showing the surface density of stellar particles
  (top), gas particles (middle) and SFR (proportional to
  UV luminosity, bottom). Scales for each are shown in the left panel, the color
  bar for the stellar particle panels is converted to projected K-band surface
  brightness in units of mag arcsec$^{-2}$, assuming a constant mass to light
  ratio of 2 M$_{\odot}/$L$_{\odot}$. If averaged radially the outer regions
  have a surface brightness of $\sim$ 25 mag arcsec$^{-2}$. SFR is proportional to UV
  luminosity, so the SFR density range corresponds to a
  range in UV luminosity range of  $7\times 10^{21}$ to $7\times 10^{29}$ ergs
  s$^{-1}$ Hz$^{-1}$ kpc$^{-2}$. The panels are 100 kpc wide.  These images
  illustrate that fiducial star formation laws create XUV
  morphology; notice how the star formation follows the peaks in the gas density
  distribution. The optical emission in the outer regions of the galaxy is
  near or below the surface brightness detection limit for most
  instruments. However, SFRs seen in the bottom images are detectable, and are similar to those seen in XUV disks (T07,
  B07). If this simulation were observed, 
  spiral structure in the outer disk would be apparent in UV
  emission without accompanying K-band emission, as in Type 1 XUV disks. }
\label{fig:sequence}
\end{figure}

The time evolution of our simulated galaxy is shown in Figure~\ref{fig:sequence}. The top
row shows the stellar surface density, the middle row shows
gas surface density, and the bottom row shows the instantaneous SFR
surface density, a rough indicator of what is observed in the UV. Since the instantaneous
SFR is shown instead of SFR over the last 5-200
Myr, this is a lower limit on the UV emission.

Figure~\ref{fig:sequence} demonstrates that spiral structure from the inner
disk propagates to the edge of the extended gas disk. In the outer disk, peaks of the
spiral waves that are above the density threshold form stars
and the
troughs that are below the density threshold do not. This gives
filamentary star formation in the outer disk, as seen in Type I XUV disks
(T07). Since this is an
isolated disk, it shows very
regular spiral structure, including ring features seen in the
second panel of Figure~\ref{fig:sequence}. In reality, peaks of spiral structure in disks will be influenced by interactions with low mass satellites and
halo substructure, in addition to the clumpiness of the ISM. The star formation in this model is not 
sufficient 
to create a high surface brightness component of stars in the optical, even after
4 Gyr. However, very low surface brightness populations of stars accompany the
XUV features. Low surface brightness populations of stars are known to exist in
outer parts of disks accompanying UV sources \citep{Zaritsky-Christlein-2007,
  Gildepaz-et-al-2007}. 

\section{Discussion} \label{sec:discuss}

\subsection{Radial profiles}

In the example presented here, 
we chose a constant density radial profile for the extended gas
disk. We ran an additional simulation with only an exponential gas disk of the same
scale length as the stellar disk and found
that while star formation took place near or slightly beyond the optical
radius, the average density of gas at several optical radii is too low to be
locally above the star formation threshold, so stars did not form there.
The choice of a constant density extended disk was motivated by examples of Type I XUV disks that
have fairly constant density extended H\,I disks \citep[e.g. NGC 4625, NGC
  5055, M83;][]{Bush-Wilcots-2004, Thilker-et-al-2005, Bosma-1981}.
Our simulation is
not intended to reproduce all XUV galaxies, but to illustrate that outer disk
star formation of the observed morphology is consistent with fiducial star
formation laws.

The gas disk we have used for this galaxy model is relatively 
extended ($\sim
45$ kpc radius), to illustrate that the size of the disk is
not an impediment to spiral structure propagating to its edge. The important
quantity for comparing to observations is the ratio of the sizes of the
optical and gas disks. Assuming that
about 2.5 scale lengths of the galaxy would be easily imaged in
the optical, the ratio of the gas to optical disk radii is about 5 for this simulation, which is 
realistic when compared to XUV disks, for example, M\,83 (T07).

\subsection{Longevity} \label{sec:longevity}

We expect XUV behavior to persist as long as there is
spiral structure in the outer disk and the peaks in the density distribution exceed the
threshold.
Some XUV emission persists throughout the
$\sim$ 4 Gyr duration of our simulation. The gas density in the
outer regions is depleted by less than 1\% during the simulation and therefore
XUV behavior will continue for the lifetime of the galaxy, as
long as this gas is not removed by another process (such as ram pressure
stripping or tidal interactions) and spiral perturbations continue.  Although XUV emission is
long-lived, its appearance is time-dependent because it follows
the spiral perturbations. For example, owing to strong spiral behavior at
1.714 Gyr, as shown in Figure~\ref{fig:sequence} and
Figure~\ref{fig:HIprofile}, we see rings of UV emission in the outer
disk. At 3.857 Gyr, there is weak spiral structure in the outer disk, leading to
only a few spots of UV emission.

In our simulation, the development of spiral structure is seeded by
numerical noise arising from the discretized dark matter halo \citep{Hernquist-1993}.
However, the simulation of XUV disks does not require a
particular mechanism for exciting spiral structure.
If there is spiral structure
in inner disks, as observed, our simulation shows that it easily
propagates to the edge of the outer disk and that UV emission tracing that spiral
structure can occur. 

\subsection{The Kennicutt-Schmidt Relation} \label{subsec:kennlaw}

\begin{figure}
\plotone{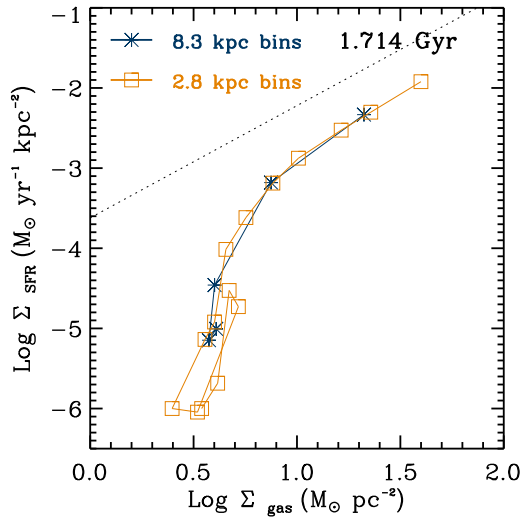}
\caption[]{Kennicutt-Schmidt relation plotted for one snapshot of our
  simulation. The gas density and SFR density are averaged in
  rings at two different bin widths out to 42 kpc. Note that annuli with a
  value of 10$^{-6}$M$_{\odot}$/pc$^{-2}$ are upper limits on the star formation
  rate.}
\label{fig:kennplot}
\end{figure}

B07 average the gas density and SFR
density in elliptical annuli of 43 nearby galaxies ($\sim$14 of which are XUV disks)
to determine whether the
Kennicutt-Schmidt relation holds when including UV emission. They find that
these galaxies fall on the Kennicutt-Schmidt relation and extend to surface
densities as low as 0.2 M$_{\odot}$ pc $^{-2}$, but with
considerable scatter. Because our
simulation has a constant density outer gas density profile at 3.5 M$_{\odot}$ pc $^{-2}$,
moderated only by spiral density waves, we do not probe star formation in
such low surface density gas.  However, in Figure~\ref{fig:kennplot} we present a 
similar analysis of our simulation,
binning one snapshot in two
bin widths, 2.8 kpc and 8.3 kpc.

It should be noted that the simulated star formation, shown in 
Figure~\ref{fig:kennplot}, is normalized slightly below the empirical Kennicutt-Schmidt law (dotted line)
so that the total SFR is approximately that of the Milky Way; however, this
will only affect the vertical normalization and not the shape within this plot.
In any case, with either the binning, our
simulation shows evidence of a cutoff, but the behavior can be surprisingly
varied. When binned
finely, bins that lie between
spiral density waves and therefore below the density threshold, fall
off the Kennicutt-Schmidt relation, so bins near the threshold density show
a wide variation in SFR. 
However, in coarser bins, this effect is smoothed as high and low density regions are included in the same annulus. Depending on
the morphology of the spiral density perturbation and the binning chosen, the
variation in SFR near the threshold density can be smoothed to the point where
these features appear to be scatter around a Kennicutt-Schmidt relation. We conclude that data
in a Kennicutt-Schmidt plot can be influenced by such factors as the binning
in addition to the underlying star formation process and need to be interpreted
carefully. The Kennicutt-Schmidt plot behavior is likely to depend sensitively
on the
gas distribution and more simulations are needed to determine whether
the B07 results can be reproduced with fiducial star
formation laws.

\section{Conclusions} \label{sec:conc} 

We show that features similar to those seen in Type I XUV disks can be
produced with fiducial star formation prescriptions, including a star
formation threshold density, if there is gas present at large radii near the
threshold density.  This is a clear illustration of the fact that a
star formation threshold density does not imply the existence of a star formation threshold radius beyond which star
formation ceases. As proposed in earlier studies and shown by
\citet{Elmegreen-Hunter-2006}, local over-densities, caused by spiral
density waves in our simulation, allow star formation in gas that, when
circularly averaged, is below or near the classical star formation threshold
density.

It is possible that star formation will differ between
inner and outer disks.  This could arise owing to different
properties of the ISM in the inner and outer regions
of disks, from, for example, a different ionized fraction owing to
various levels of self-shielding against an ionizing UV background.
Currently, our simulations do not have the resolution to probe the
detailed physics of star formation.  
Here, we simply show that any simulation whose star
formation prescription follows a Kennicutt-Schmidt relation with a
star formation threshold density and has a high enough gas density for
local over-densities to exceed the star formation threshold density in
the outer disk will exhibit UV emission in the outer disk.

Understanding the mechanism for generating XUV disks has promising
implications for probing galaxy structure in the outer parts of
galaxies.  If the distribution of H\,I can be determined based on the
observed UV emission, this can be used to infer
the extended H\,I properties of disks.
Outer disk H\,I and its star formation
history could also constrain recent gas accretion in galaxies and the
formation theory of disks. For these reasons, the
generation of XUV disks warrants further exploration.  
Simulations having different extended gas profiles (e.g. exponential,
$1/R$) at various gas densities are needed to test whether we can
create the range of properties seen in XUV disks, including Type
II XUV disks. Upcoming extragalactic H\,I surveys such as THINGs
\citep{Walter-et-al-2005} and ALFALFA \citep{Giovanelli-et-al-2005} will provide 
better statistics on the types of H\,I profiles seen on disk
galaxies and how they correlate with XUV disk behavior. The role of interactions, particularly flybys, in
exciting spiral structure and affecting the morphology of XUV disks
also needs to be explored.  Finally, the application of a radiative
transfer code to our SPH results to compare colors of these clumps to
samples such as T07 and \citet{Zaritsky-Christlein-2007}
would provide interesting constraints on star formation
such as the initial mass function in outer disks.

\acknowledgements
We appreciate the helpful comments of the referee, Joop Schaye. We thank Phil Hopkins, Dusan Keres, Zhong Wang and
Jacqueline van Gorkom for useful conversations in the preparing this
work.  Support for TJC was provided by the W.M. Keck Foundation.

%\bibliography{xuv_v4.bib}

\begin{thebibliography}{34}
\expandafter\ifx\csname natexlab\endcsname\relax\def\natexlab#1{#1}\fi

\bibitem[{{Andersen} {et~al.}(2006)}]{Anderson-et-al-2006}
{Andersen}, D.~R. {et~al.} 2006, \apjs, 166, 505

\bibitem[{{Boissier} {et~al.}(2007)}]{Boissier-et-al-2007}
{Boissier}, S. {et~al.} 2007, \apjs, 173, 524

\bibitem[{{Bosma}(1981)}]{Bosma-1981}
{Bosma}, A. 1981, \aj, 86, 1791

\bibitem[{{Brand} {et~al.}(2001){Brand}, {Wouterloot}, {Rudolph}, \& {de
  Geus}}]{Brand-et-al-2001}
{Brand}, J., {Wouterloot}, J.~G.~A., {Rudolph}, A.~L., \& {de Geus}, E.~J.
  2001, \aap, 377, 644

\bibitem[{{Bush} \& {Wilcots}(2004)}]{Bush-Wilcots-2004}
{Bush}, S.~J. \& {Wilcots}, E.~M. 2004, \aj, 128, 2789

\bibitem[{{Christlein} \& {Zaritsky}(2008)}]{Christlein-Zaritsky-2008}
{Christlein}, D. \& {Zaritsky}, D. 2008, ArXiv e-prints, 803

\bibitem[{{Dong} {et~al.}(2008){Dong}, {Calzetti}, {Regan}, {Thilker},
  {Bianchi}, {Meurer}, \& {Walter}}]{Dong-et-al-2008}
{Dong}, H., {Calzetti}, D., {Regan}, M., {Thilker}, D., {Bianchi}, L.,
  {Meurer}, G.~R., \& {Walter}, F. 2008, ArXiv e-prints, 804

\bibitem[{{Elmegreen} \& {Hunter}(2006)}]{Elmegreen-Hunter-2006}
{Elmegreen}, B.~G. \& {Hunter}, D.~A. 2006, \apj, 636, 712

\bibitem[{{Elmegreen} \& {Parravano}(1994)}]{Elmegreen-Parravano-1994}
{Elmegreen}, B.~G. \& {Parravano}, A. 1994, \apjl, 435, L121+

\bibitem[{{Ferguson} {et~al.}(1998){Ferguson}, {Wyse}, {Gallagher}, \&
  {Hunter}}]{Ferguson-et-al-1998}
{Ferguson}, A.~M.~N., {Wyse}, R.~F.~G., {Gallagher}, J.~S., \& {Hunter}, D.~A.
  1998, \apjl, 506, L19

\bibitem[{{Gil de Paz} {et~al.}(2005)}]{Gildepaz-et-al-2005}
{Gil de Paz}, A. {et~al.} 2005, \apjl, 627, L29

\bibitem[{{Gil de Paz} {et~al.}(2007)}]{Gildepaz-et-al-2007}
---. 2007, \apj, 661, 115

\bibitem[{{Giovanelli} {et~al.}(2005)}]{Giovanelli-et-al-2005}
{Giovanelli}, R. {et~al.} 2005, \aj, 130, 2598

\bibitem[{{Hernquist}(1990)}]{Hernquist-1990}
{Hernquist}, L. 1990, \apj, 356, 359

\bibitem[{{Hernquist}(1993)}]{Hernquist-1993}
---. 1993, \apjs, 86, 389

\bibitem[{{Kennicutt}(1989)}]{Kennicutt-1989}
{Kennicutt}, Jr., R.~C. 1989, \apj, 344, 685

\bibitem[{{Kennicutt}(1998)}]{Kennicutt-1998}
---. 1998, \apj, 498, 541

\bibitem[{{Martin} \& {Kennicutt}(2001)}]{Martin-Kennicutt-2001}
{Martin}, C.~L. \& {Kennicutt}, Jr., R.~C. 2001, \apj, 555, 301

\bibitem[{{Martin} {et~al.}(2005)}]{Martin-et-al-2005}
{Martin}, D.~C. {et~al.} 2005, \apjl, 619, L1

\bibitem[{{Quirk}(1972)}]{Quirk-1972}
{Quirk}, W.~J. 1972, \apjl, 176, L9+

\bibitem[{{Schaye}(2004)}]{Schaye-2004}
{Schaye}, J. 2004, \apj, 609, 667

\bibitem[{{Schmidt}(1959)}]{Schmidt-1959}
{Schmidt}, M. 1959, \apj, 129, 243

\bibitem[{{Spitzer}(1968)}]{Spitzer-1968}
{Spitzer}, Jr., L. 1968, (New York, NY (USA): John Wiley {\&} Sons), No.~28,
  XIII + 262 p., 28

\bibitem[{{Springel}(2005)}]{Springel-2005}
{Springel}, V. 2005, \mnras, 364, 1105

\bibitem[{{Springel} {et~al.}(2005){Springel}, {Di Matteo}, \&
  {Hernquist}}]{Springel-Dimatteo-Hernquist-2005}
{Springel}, V., {Di Matteo}, T., \& {Hernquist}, L. 2005, \mnras, 361, 776

\bibitem[{{Springel} \& {Hernquist}(2003)}]{Springel-Hernquist-2003}
{Springel}, V. \& {Hernquist}, L. 2003, \mnras, 339, 289

\bibitem[{{Thilker} {et~al.}(2005)}]{Thilker-et-al-2005}
{Thilker}, D.~A. {et~al.} 2005, \apjl, 619, L79

\bibitem[{{Thilker} {et~al.}(2007)}]{Thilker-et-al-2007}
---. 2007, \apjs, 173, 538

\bibitem[{{Toomre}(1964)}]{Toomre-1964}
{Toomre}, A. 1964, \apj, 139, 1217

\bibitem[{{van der Kruit} \& {Shostak}(1984)}]{vanderKruit-Shostak-1984}
{van der Kruit}, P.~C. \& {Shostak}, G.~S. 1984, \aap, 134, 258

\bibitem[{{van Zee} {et~al.}(1997){van Zee}, {Haynes}, {Salzer}, \&
  {Broeils}}]{vanZee-et-al-1997}
{van Zee}, L., {Haynes}, M.~P., {Salzer}, J.~J., \& {Broeils}, A.~H. 1997, \aj,
  113, 1618

\bibitem[{{Walter} {et~al.}(2005){Walter}, {Brinks}, {de Blok}, {Thornley}, \&
  {Kennicutt}}]{Walter-et-al-2005}
{Walter}, F., {Brinks}, E., {de Blok}, W.~J.~G., {Thornley}, M.~D., \&
  {Kennicutt}, R.~C. 2005, in Astronomical Society of the Pacific Conference
  Series, Vol. 331, Extra-Planar Gas, ed. R.~{Braun}, 269--+

\bibitem[{{Wong} \& {Blitz}(2002)}]{Wong-Blitz-2002}
{Wong}, T. \& {Blitz}, L. 2002, \apj, 569, 157

\bibitem[{{Zaritsky} \& {Christlein}(2007)}]{Zaritsky-Christlein-2007}
{Zaritsky}, D. \& {Christlein}, D. 2007, \aj, 134, 135

\end{thebibliography}

\clearpage

\end{document}